\documentclass[A4paper,11pt]{article}
\usepackage[english]{babel}
\usepackage{hyperref}
\usepackage[pdftex]{graphicx}

\begin{document}
\pagenumbering{arabic}
\titlepage {
\title{Performance of the PADME calorimeter prototype at the DA$\Phi$NE BTF}

\author{M.~Raggi$^{c,d}, $
V.~Kozhuharov$^{a,b}, $
P.~Valente$^d$, 
F.~Ferrarotto$^d$, 
E.~Leonardi$^d$,\\ 
G.~Organtini$^{c,d}$,
L.~Tsankov$^b$, 
G.~Georgiev$^b$, 
J.~Alexander$^e$, \\
B.~Buonomo$^a$, 
C.~Di Giulio$^a$, 
L.~Foggetta$^a$,
G.~Piperno$^a$\\ \\ 
$^a$ Laboratori Nazionali di Frascati, 00044 Frascati (RM), Italy\\
$^b$ Faculty of Physics, University of Sofia ``St. Kl. Ohridski'', 5 J. Bourchier\\ Blvd., 1164 Sofia, Bulgaria\\
$^c$ Sapienza Universit\`a di Roma, Piazzale Aldo Moro 5, 00185 Rome, Italy\\
$^d$ INFN sezione di Roma, Piazzale Aldo Moro 5, 00185 Rome, Italy\\
$^e$ Cornell University, Ithaca, NY 14853, USA\\
}
\maketitle

\abstract{
The PADME experiment at the DA$\Phi$NE Beam-Test Facility (BTF) aims at searching for invisible decays of the dark photon by measuring
 the final state missing mass in the process $e^+e^- \to \gamma+ A'$, with $A'$ undetected.
The measurement requires the determination of the 4-momentum of the recoil photon, performed using a homogeneous,
 highly segmented BGO crystals calorimeter. We report the results of the test of a 5$\times$5 crystals prototype 
 performed with an electron beam at the BTF in July 2016. }
}

\section{Introduction}

A possible solution to the dark matter problem postulates that dark matter interacts with standard model particles through a new 
force mediated by a ``portal". If the new force has a U(1) gauge structure, the ``portal" is a massive photon-like vector particle, 
called Dark Photon or $A'$.
In the most general scenario the existance of dark sector particles with a mass below that of $A'$ is not excluded: in this case so-called ``invisible'' decays of the $A'$ are allowed.
Moreover, given the small coupling of the $A'$ to visible SM particles, which makes the visible rates
suppressed by $\epsilon^2$ ($\epsilon$ being the reduction factor of the coupling of the dark photon with respect to the electromagnetic one), 
it is not hard to realize a situation where the invisible decays dominate.
There are several studies on the searches of a $A'$ decaying into
dark sector particles ($\chi$), recently summarized in \cite{Raggi:2015yfk}\cite{Alexander:2016aln}. 

The aim of the PADME experiment is to detect the non Standard Model process $e^+e^- \to \gamma ~ + ~ A'$, $A'$ undetected,
by measuring the missing mass in the final state \cite{Raggi:2014zpa}\cite{Raggi:2015gza}, using 550 MeV positrons from the improved
Beam-Test Facility (BTF) of the Frascati LINAC \cite{Valente:2016tom}. 
The experiment is composed of a thin (100 $\mu$m in the baseline design) active diamond target, to
measure the average position and the intensity of the positrons during a single beam pulse, a set of charged particle veto detectors immersed
in the field of a dipole magnet, to detect the positron losing their energy due to Bremsstrahlung radiation,
and a calorimeter, to measure/veto final state photons. The apparatus is inserted into a
vacuum chamber, to minimize unwanted interactions of primary and secondary particles that might generate extra photons.
The rate in the central part of the calorimeter is too high due to Bremsstrahlung photons. For this reason the calorimeter has a central hole covered
by a faster photon detector, the Small Angle Calorimeter (SAC). The maximum repetition rate of the beam pulses is 49 Hz.
In the following sections we describe the PADME BGO calorimeter, the $5\times5$ cells prototype being tested, and the results on the 
prototype performance obtained during a test with electrons at BTF in July 2016.

\section{The PADME calorimeter}

\begin{figure}[htb]
\begin{center}
\includegraphics[width=6.5cm]{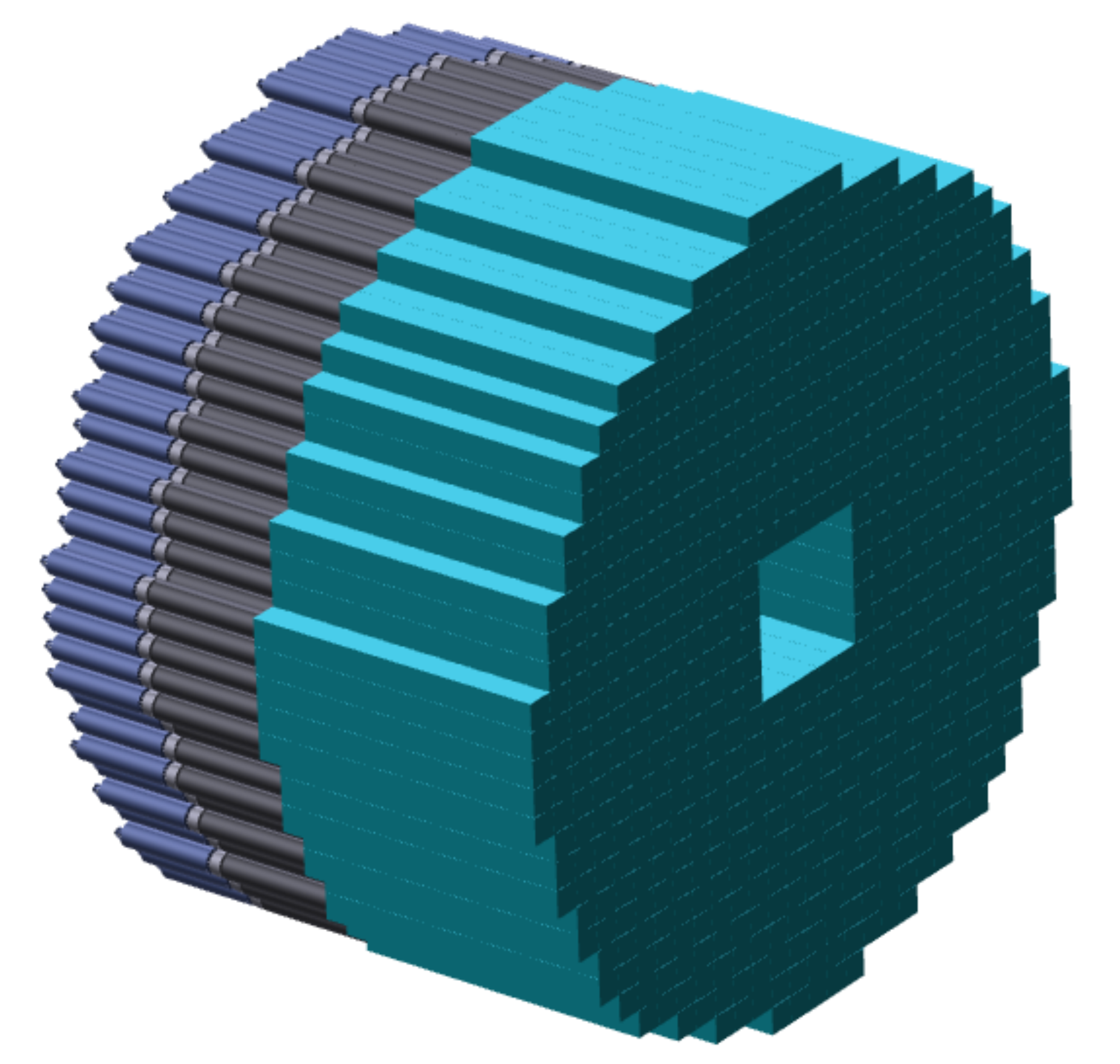}
\end{center}
\caption{The PADME BGO crystal calorimeter.}
\label{fig:calo3d}
\end{figure}

The PADME calorimeter is a homogeneous crystal calorimeter with an approximately cylindrical shape,
with a diameter of $\sim$600 mm, depth of 230 mm, and with a central 100$\times$100 mm$^2$ square hole (see Fig.\ref{fig:calo3d}).
The active volume will be composed by 616 21$\times$21$\times$230 mm$^3$ BGO crystals, obtained by machining 
the crystals recovered from one of the end-caps of the electromagnetic calorimeter of the dismantled L3 experiment at LEP\cite{Adeva:1990}.
According to the tests performed by the L3 collaboration \cite{Karyotakis:1995ki}, the expected energy resolution lies in the interval $(1-2)\%/\sqrt E$ for $<1$ GeV electrons and photons. 

Early tests aimed at evaluating the best readout technology showed that avalanche photodiodes (APDs), even with a (relatively large) area of 10$\times$10 mm$^2$, have a gain, and consequently a total collected charge, that is insufficient to perform a 
high resolution energy measurement in the energy interval relevant to PADME, that is from a few to a few hundred MeV.
The readout system will therefore be based on 19mm diameter photo-multiplier tubes.

\section{The Frascati Beam test Facility}
The Beam Test Facility (BTF) of the DA$\Phi$NE LINAC \cite{Ghigo:2003gy} at the Frascati National 
Laboratory of INFN (LNF), is a dedicated
beam line which delivers electrons or positrons in 
a dedicated experimental hall. The beam can be used for detector testing purposes
and/or studying physics phenomena at the energy scale of O(100 MeV). 
BTF can deliver up to 49 pulses of $e^+/e^-$ per second, with energy 
in the range starting at $\approx$50 MeV up to 750 MeV for $e^-$ and 550 MeV for $e^+$.

An important feature of the facility is the possibility to vary the 
beam intensity from $\sim 10^9$ particles down to a single particle per pulse. 
The latter is extremely important for detector efficiency and energy resolution 
measurements. The beam pulse duration can be changed in the range 1.5-40 ns, by
changing the number of 2856 MHz bunches from the LINAC, even though it is fixed
to 10 ns during the operation of the Frascati DAFNE collider. This was also the case 
during the PADME calorimeter prototype test.

During the July beam-test the BTF was operated in electron mode and the average number of 
particles per single bunch was kept close to one, in order to study the
response of the PADME calorimeter prototype. 
Two different beam setups  with energies of 250 and 450 MeV were used.  
The typical size of the beam spot was kept by the BTF optics below 1 cm$^2$ RMS
while the energy spread of the beam was estimated at the level of 1\%.

\section{The PADME calorimeter prototype}
The prototype was composed of 25 BGO crystals  arranged in a 5$\times$5 matrix. 
The geometry was obtained by machining original L3 crystals to get a parallelepiped of 20$\times$20$\times$220 mm$^3$, 
very close to the final dimensions for the experiment.
The crystals were wrapped with 
teflon sheets and 
the scintillation light was detected by 19 mm diameter photo-multipliers
(15 mm diameter active area) by HZC Photonics\footnote{\path{http://hzcphotonics.com/}}, model XP1912, coupled to the crystals using optical grease.

The prototype was placed on the BTF remotely movable table, and was 
adjusted so that the beam impinged onto the central crystal of the matrix. 
The  photo-tubes were operated at $\sim$ 1100 V, corresponding to an equalized gain of
$\sim5\times10^5$, 
according to the HZC Photonics specifications.

\begin{figure}[htb]
\begin{center}
\includegraphics[width=6cm,angle=-90]{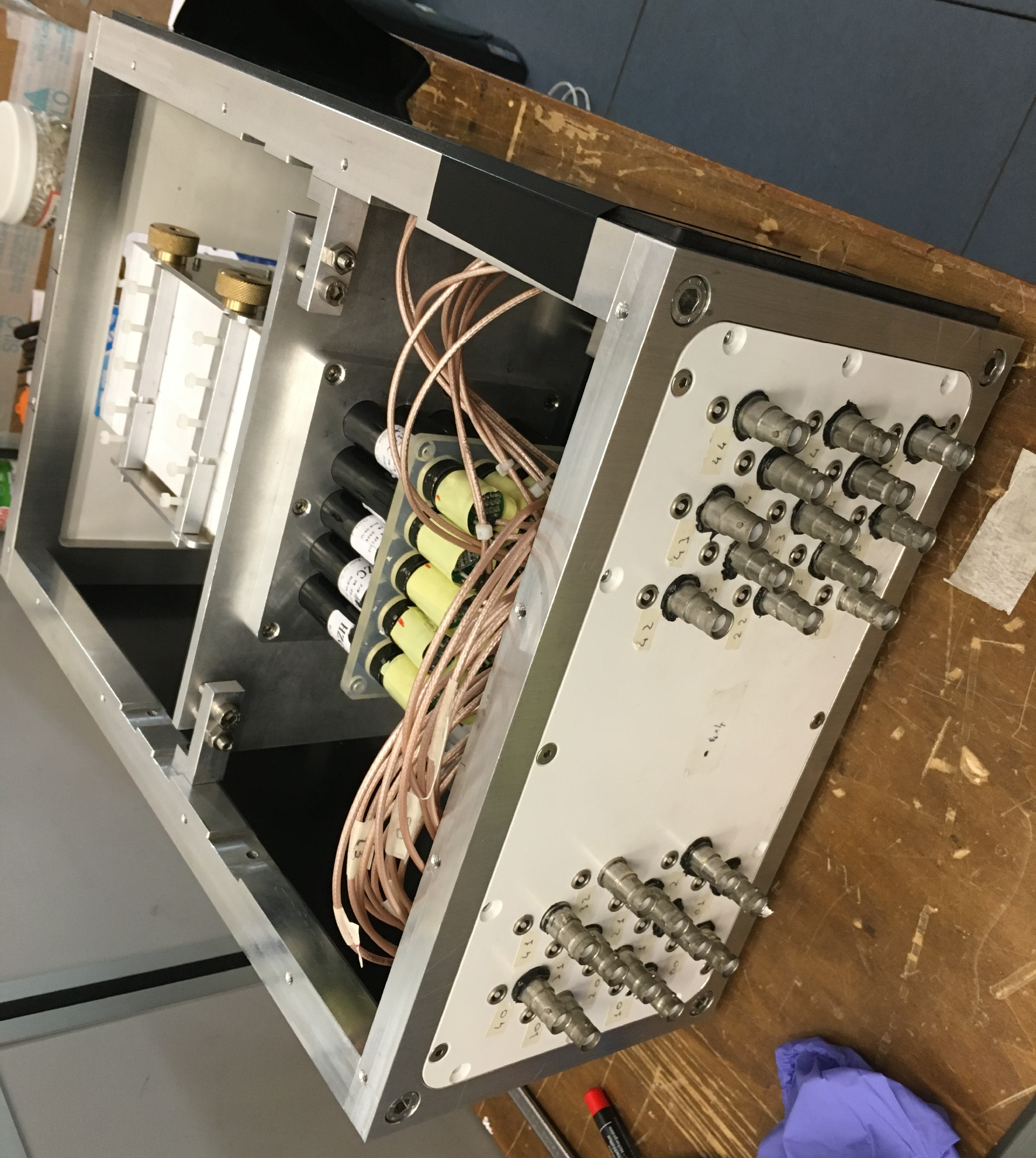}
\end{center}
\caption{The PADME 5$\times$5 BGO crystal prototype.}
\label{fig:btf-setup}
\end{figure}

The 25 channels of the prototype were fed into a CAEN V1742 high-speed digitizer\cite{bib:caen-v1742},
based on the DRS4 chip, set to a sampling speed of 1 GS/s (1 ns/sample). 
The digitizer was operated in sampling mode providing 12 bit measurement of the input amplitude for
the 1024 samples, corresponding to an integration window of $\approx$ 1 $\mu$s. 
The trigger was based on an external NIM signal from the BTF timing system, which  
allowed to record the waveforms of all the readout channels for every single pulse. The timing with
respect to the actual arrival of electrons at the BTF beam exit was adjusted by means of the BTF
programmable digital delay.
The data were transferred to a readout PC through optical fibers by a dedicate control program and
stored in binary format for further analysis. 

The presented results are based on the data sample collected during two week test
run at BTF in July 2016.

\section{Charge reconstruction}

The offline data contained the recorded waveforms in a window of
1024 ns (1024 amplitude measurements every 1 ns for each event). 
The typical BGO signal has a duration of $\sim$1 $\mu$s due to the 300 ns decay constant of the scintillation light. 
To integrate as much signal as possible, the BTF trigger was set to about 100 ns prior to the electrons arrival time. 
For each recorded event (corresponding to a single beam pulse)
the individual PMT charges were obtained by integrating the 
recorded waveforms after pedestal subtraction.
The pedestal was determined on event by event basis using the average of the first 100 samples, i.e. the ones preceding the 
start of the scintillation signal, in order to keep within the data acquisition window the largest possible fraction of the signal.
The total charge was calculated summing the signals from all the photo-tubes in which the value of the charge was greater than the pedestal.

\begin{figure}[htb]
\begin{center}
\includegraphics[width=6.cm, angle=-90]{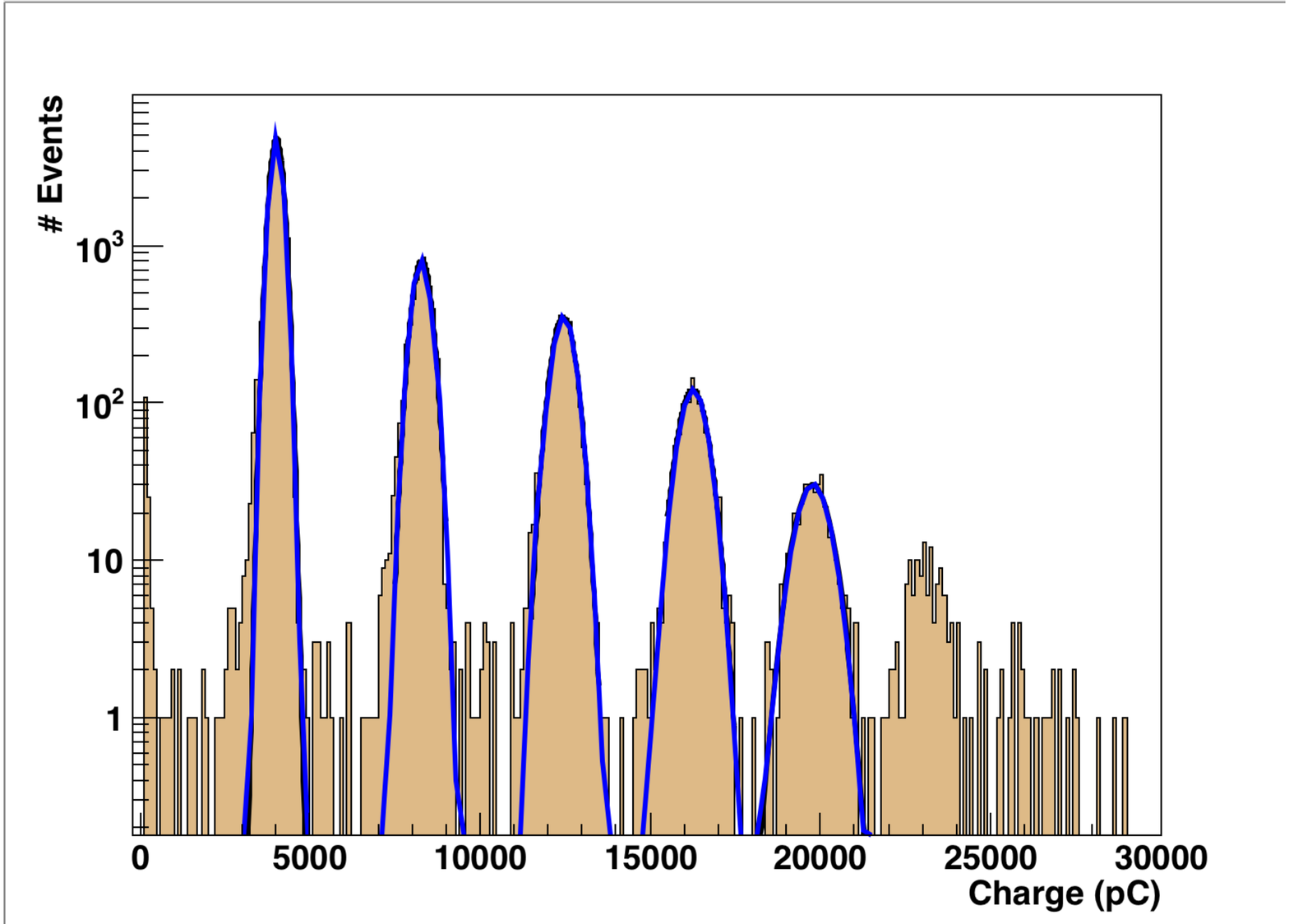}   
\end{center}
\caption{Total reconstructed charge in the prototype for 250 MeV electrons. 
The distribution was approximated with the sum of five gaussians, 
with parameters determined from a free fit.}
\label{fig:tot-charge}
\end{figure}
The events were selected requiring that the number of crystals above a given threshold was greater than 10, 
and that the reconstructed electron impinging position of the electron laid in a circle of 5 mm radius around the center 
of the prototype. The threshold has been set to ....

The reconstructed charge distribution for the 250 MeV data sample 
is shown in Fig. \ref{fig:tot-charge}. The individual 
peaks correspond to one, two, etc. electrons impinging the 
detector. 
A fit on the collected charge with the sum of five Gaussians was performed:
\begin{equation}
 N(x) = \sum_{i=1}^n N_i \times \exp\left( \frac{-(x - q_i)^2}{2\sigma_i^2} \right),
 \label{eqn:gausfit}
\end{equation}
where $q_i$ is the charge  corresponding to $i$ electrons 
in the detector, $\sigma_i$ is the charge resolution and $N_i$ are normalization factors.
The fit was performed on the first 5 Gaussians ($n=5$).
All 15 parameters were left free, resulting in the best fit shown with a blue line 
in Fig.\ref{fig:tot-charge}. 

The center of each of the Gaussians 
allows the extraction of the relation between the collected charge 
and the deposited energy. 

\section{Detector performance}

\begin{figure}[htb]
\begin{center}
\includegraphics[width=7cm]{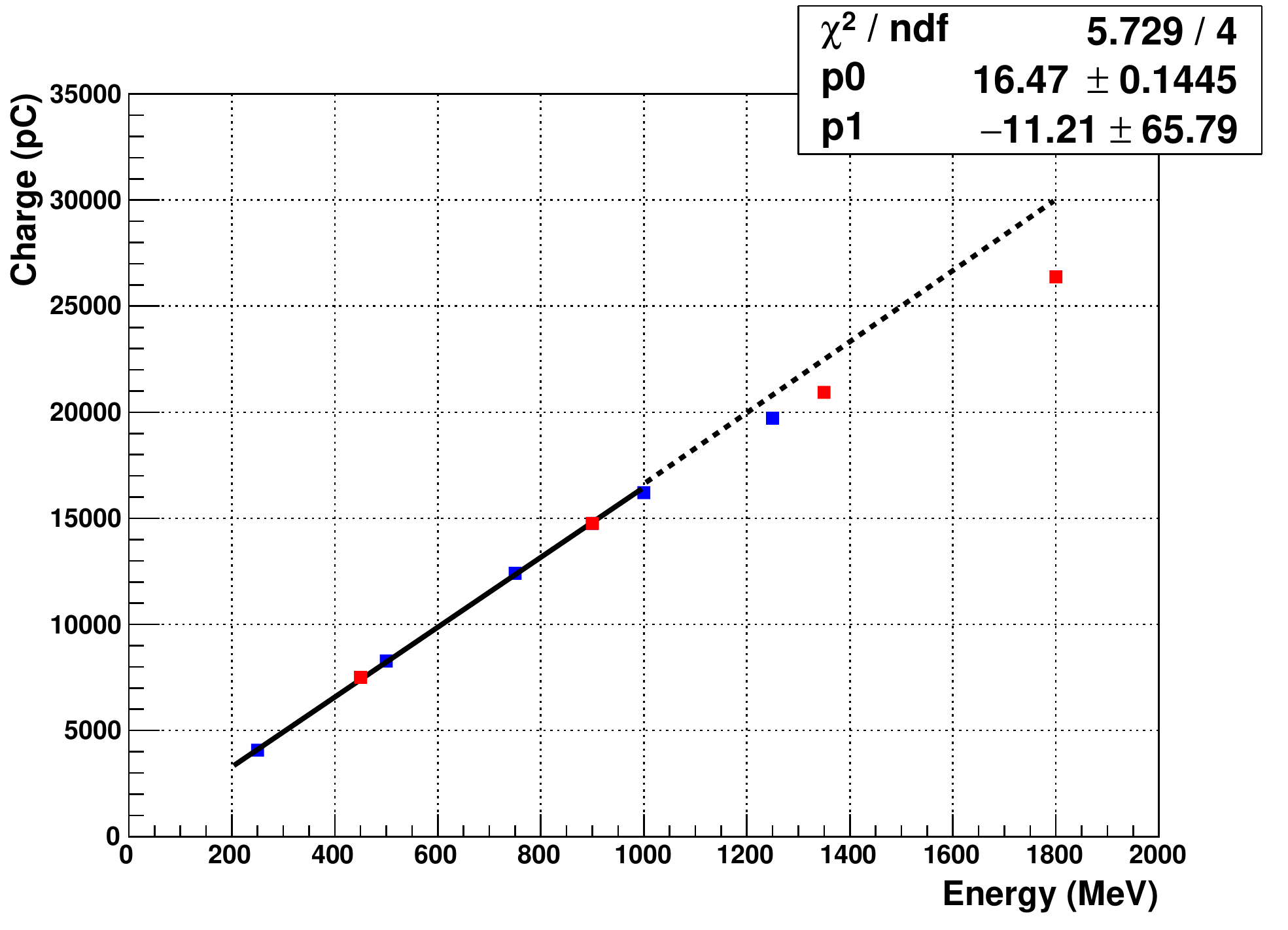}   
\end{center}
\caption{Relation between the reconstructed charge in the PADME calorimeter prototype and the deposited energy.
The linear fit to the points (black line) and its extrapolation to higher energies (dashed line) are also shown.}
\label{fig:Ecal-lin}
\end{figure}

The relation (shown in Fig.\ref{fig:Ecal-lin})
was verified for energies up to 1.8 GeV and is linear within 2\% up to $n=4$ electrons, corresponding to a total energy of 1 GeV, with a 
slope of $\sim$16.5 pC/MeV (black line in Fig.\ref{fig:Ecal-lin}).
For higher energies the 1 V maximum dynamical range of the V1742 digitizer \cite{bib:caen-v1742}
is exceeded, so that the charge measurement obtained by simply summing the sampled voltages starts to be biased, 
resulting in a non linear behaviour, as shown by the dashed line in Fig.\ref{fig:Ecal-lin}, representing the extrapolation of the measured linearity.
   
\begin{figure}[htb]
\begin{center}  
\includegraphics[width=8.5cm]{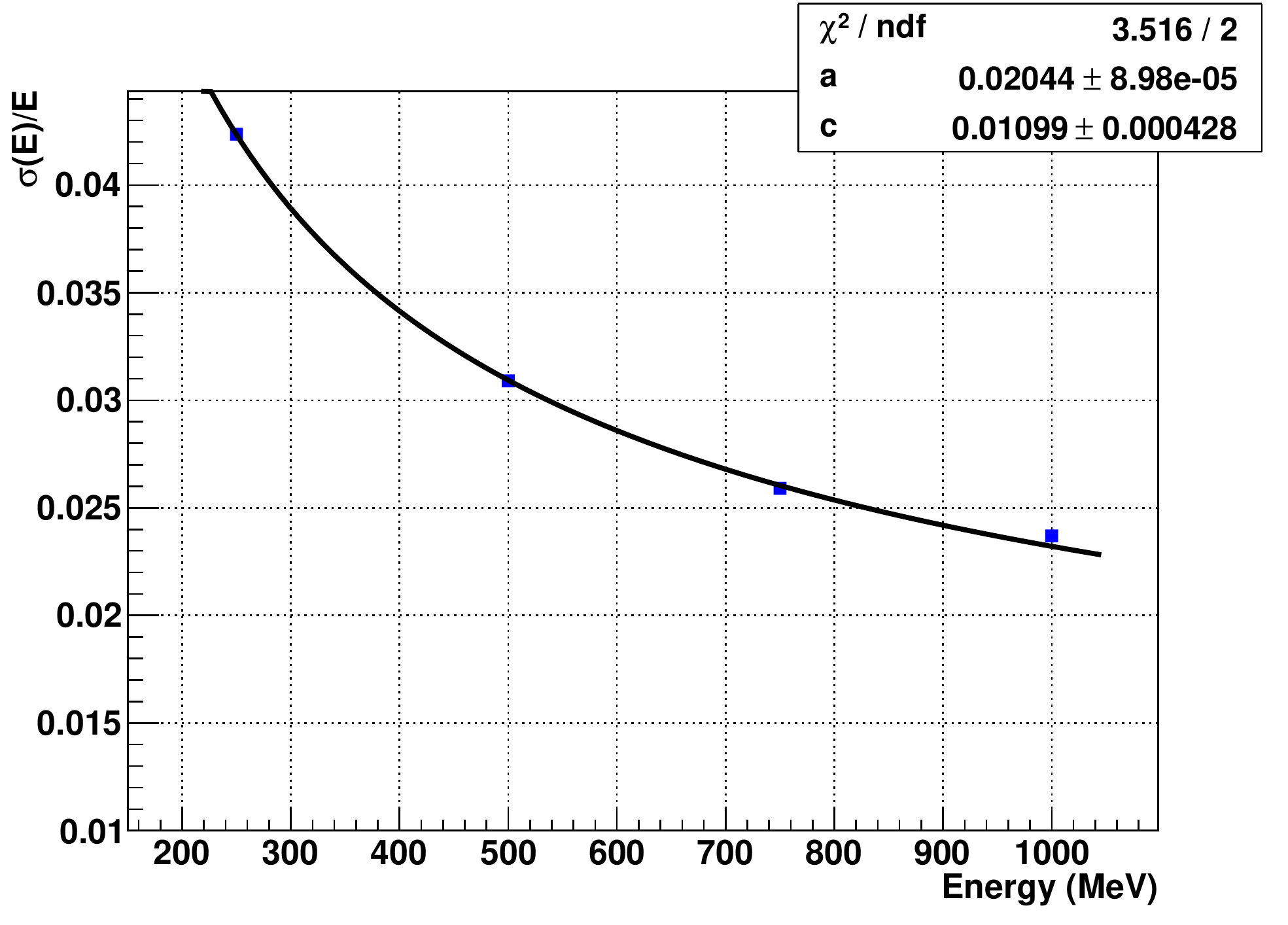}   
\end{center}
\caption{Energy resolution of the PADME calorimeter prototype as a function of the deposited energy (blue points)
approximated by the resolution function.}
\label{fig:Calo-Eres}
\end{figure}

In the non linear regime, the precision on the charge reconstruction is dominated by saturation effects. For this reason, only energies 
below 1 GeV are included in the energy resolution measurements. As in the PADME experiment the photon energy
is expected to be less than 550 MeV, this energy range is satisfactory. 
The dependence of the width of the Gaussians 
normalized to the corresponding peak values (i.e. the energy resolution, $\sigma(E)/E$) 
as a function of the total energy for the 250 MeV data sample is shown in Fig.\ref{fig:Calo-Eres}. 
This dependence is fitted with the relation
\begin{equation}
\sigma(E)/E = a/\sqrt{E} \oplus c
\label{eq:energy-res}  
\end{equation}
where the energy is measured in MeV.
The result of the fit, shown in Fig.\ref{fig:Calo-Eres} with a black line, determined the two free parameters
to be $a =2.0\%$ and $c = 1.1\%$. 

\section{Comparison with 450 MeV sample}

\begin{figure}[htb]
\begin{center}  
\includegraphics[width=8.5cm]{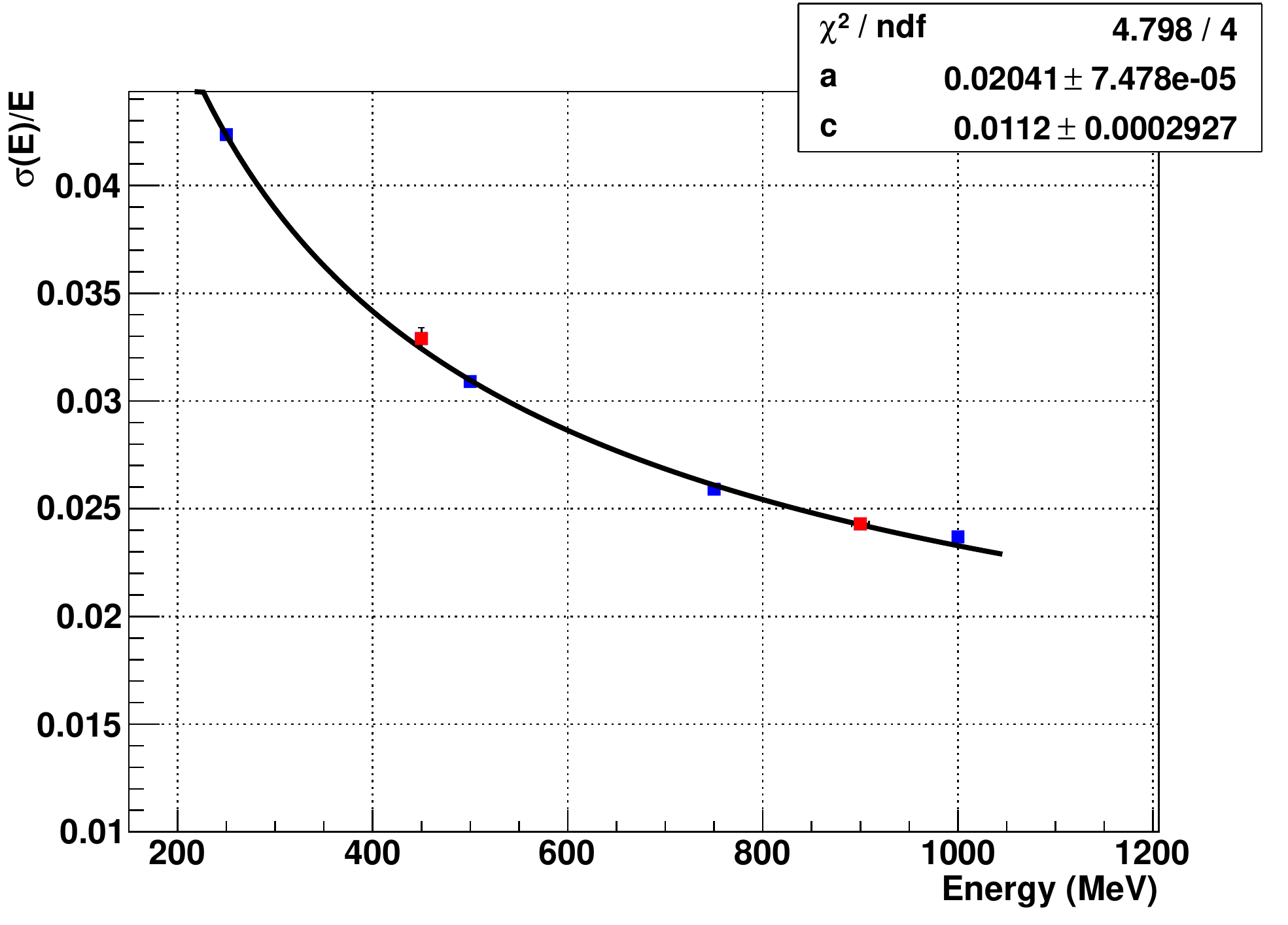}   
\end{center}
\caption{Energy resolution of the PADME calorimeter prototype as a function of the deposited energy (250 MeV $e^-$ blue points and 450 MeV $e^-$ red)
approximated with the resolution function.}
\label{fig:Calo-EresTot}
\end{figure}

Additional information on the energy resolution can be extracted from the 450 MeV electrons data sample. 
The 25 high voltage settings were kept unchanged, in order to collect a compatible data set with
the same photo-tubes gain. After reconstructing the charge and fitting with function (\ref{eqn:gausfit}), 
only the first two points fall on the straight line expected from the 250 MeV electrons linearity measurement. 
When these two points, corresponding to a total energy of 450 and 900 MeV, are added and the fit to 
function (\ref{eq:energy-res}) is repeated (see Fig.\ref{fig:Calo-EresTot}), the fit parameters are basically unchanged, pointing to a very good quality 
of the beam and reproducibility of the detector conditions.

\section{Comparison with Monte Carlo simulation}

To better understand the data collected, the result on the energy resolution has been compared with the
prediction obtained from the PADME Monte Carlo program.
The PADME experiment MC simulation implements a calibration mode, which allows to 
direct particles of a given energy to a defined part 
of the detector. Using this mode, we generated a sample of electrons with energies ranging from 
50 MeV to 1 GeV in steps of 50 MeV, pointing to the center of the active area of the calorimeter, 
i.e. far from the internal and external borders of the cylinder, with a spot of 1 cm radius and 1\% energy spread.
The reconstruction algorithm collects the energy deposits in a 5$\times$5 crystals matrix centered on the impact position.
Only crystals with a deposited energy above a given threshold were considered giving a non-zero contribution 
to the total charge. The contribution of each crystal to the total charge was then evaluated by converting 
the deposited energy to photo-electrons ($N_{p.e.}\times E(i)$) and then applying the statistical fluctuations on the number of 
photo-electrons. After multiplying it by the photo-tube gain and electron charge ($Ge$) the individual charges were scaled by 
a set of calibration constants $C(i)$. The total charge was computed according to the formula:
\begin{equation}
 Q_{tot} = \sum_{i=0}^{N_{Cry}} (N_{p.e.}E(i)Ge)\times C(i)
 \label{eqn:Qtot}
\end{equation}

This algorithm includes three free parameters that can be extracted from the collected data, and it can be tuned 
to reproduce the measured energy resolution.
The first parameter is the minimum energy deposit necessary to produce a signal above pedestal in a given crystal.
This condition corresponds to applying the zero suppression cut on the data.
Using the ratio between the peak energy and the pedestal in the data, the energy threshold was evaluated to be of the order of 2 MeV.
The number of photo-electrons ($N_{p.e.}$) dominates the energy resolution at low energy. 
From the slope of the linearity fit and the nominal photo-tube gain, we extracted a value of $N_{p.e.}\simeq 200$~p.e./MeV.
 The determination of crystal-to-crystal calibration constants was not performed during the data-taking and therefore no 
 information is available to tune the Monte Carlo.
 We decided to produce random calibration constants to be applied to single crystals to set the order of miscalibration 
that was present during the data taking.   

\begin{figure}[htb]
\begin{center}  
\includegraphics[scale=0.45]{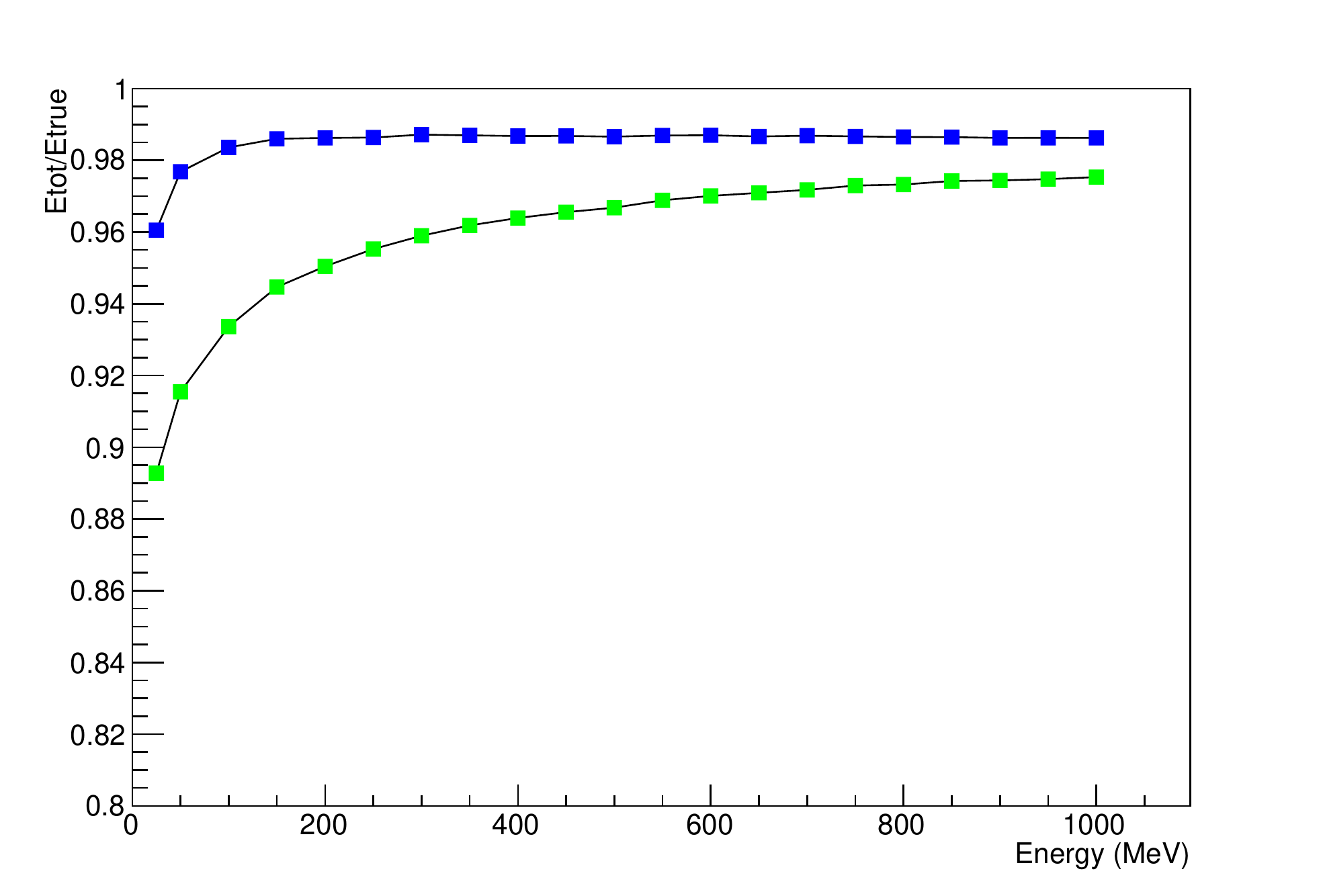}   
\end{center}
\caption{Total energy in the 5$\times$5 matrix (blue dots) and in the reconstructed cluster (green dots).}
\label{fig:CollEnergy} 
\end{figure}

\begin{figure}[htb]
\begin{center}  
\includegraphics[scale=0.45]{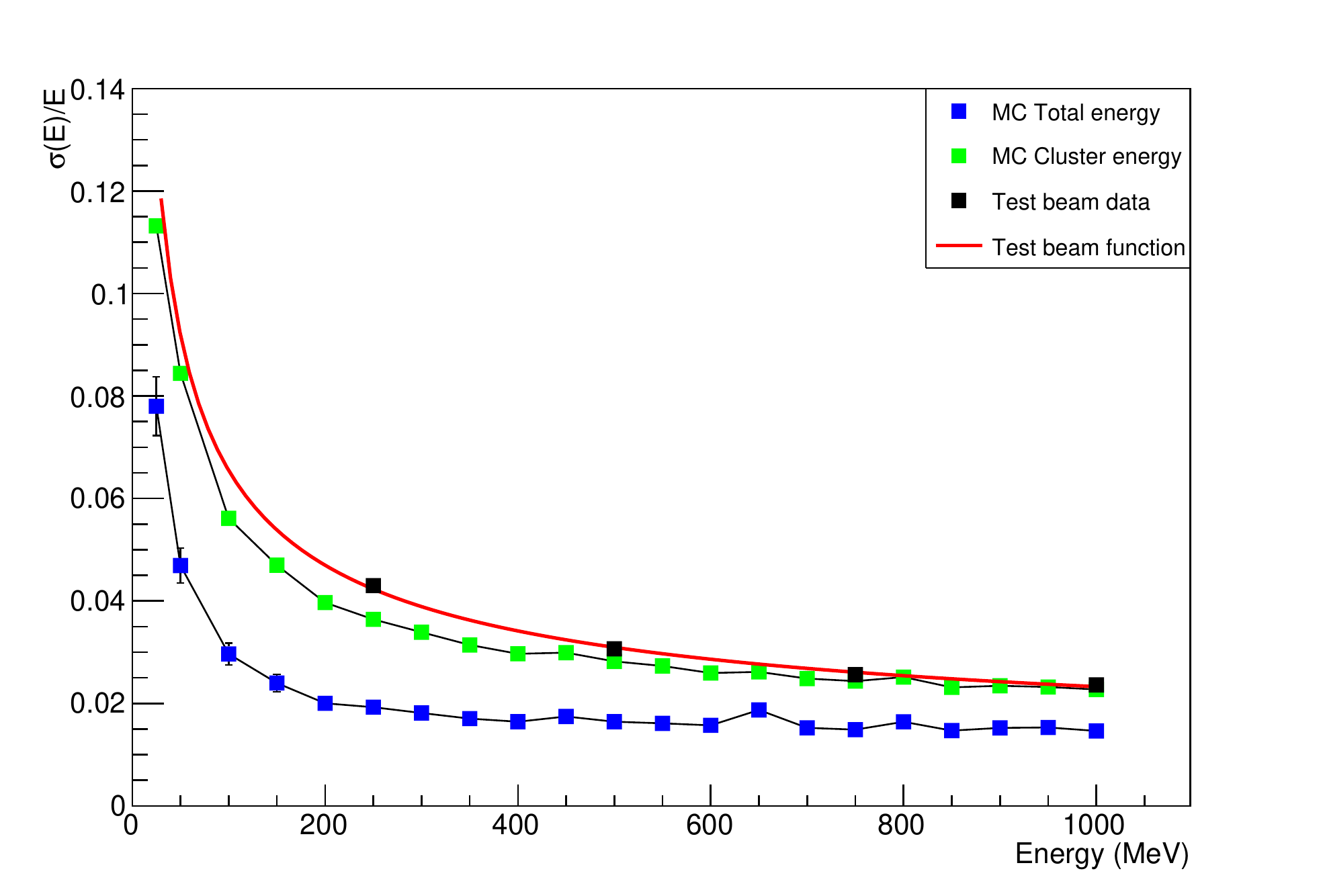}   
\end{center}
\caption{Energy resolution during the test beam (red line) compared to reconstructed energy resolution (green dots) and total 
energy resolutions (blue dots) in MC simulations. Black dots represent the resolution measured with the 250 MeV data sample.}
\label{fig:DataMC} 
\end{figure}

Fig.\ref{fig:CollEnergy} shows the total energy deposited in the 5$\times$5 crystal matrix (blue dots) 
and the corresponding reconstructed energy (green dots). The increasing difference between the two curves 
at low energy clearly shows the effect of the minimum energy threshold, which strongly enhances the fraction of energy 
lost by the reconstruction algorithm. 
This effect dominates the energy resolution at low energy and can be mitigated by increasing the photo-tubes gain, 
or by reducing the RMS of the pedestal distribution.
Fig.\ref{fig:DataMC} shows the comparison of the energy resolution measured at the beam-test (red line) 
with the Monte Carlo prediction obtained using the parameters measured from beam-test data 
(green dots). A good agreement is achieved over all the energy range, even when extrapolating the curve below the minimum measured 
energy point (black dots) of 250 MeV. The blue dots, representing the total collected energy by 
the 5$\times$5 crystal matrix, set a lower limit to the energy resolution achievable in a perfect calorimeter with a PADME-like geometry.

\section{Conclusions}

The PADME calorimeter prototype has a linear response better than 2\% for energies up to 1 GeV, with a slope of 
$\sim$16.5 pC/MeV, at a gain of $\sim5\times 10^5$.
The obtained energy resolution is
$\sigma(E)/E = 2.0\%/\sqrt{E} \oplus 1.1\%$. 
The prototype performance fulfils the PADME experiment requirements, as defined in \cite{Raggi:2014zpa}. 
The estimated number of photoelectrons/MeV produced by the prototype is $\simeq 200$~p.e./MeV.
Montecarlo studies suggest that an improvement of the resolution may come from reducing the energy threshold 
in each crystal, thus improving the fraction of energy collected in the cluster at low energy. 
To this end, a dedicated random trigger could be implemented in data taking: this would improve the knowledge 
of the pedestals and help reducing their RMS. Further improvement can come from increasing the gain 
of the prototype photomultipliers and by performing a more accurate crystal-to-crystal calibration.

\section*{Acknowledgements}

We warmly thank the BTF and LINAC teams,
for the excellent quality of the beam.
The authors are also grateful to E. Capitolo, C. Capoccia, and R. Lenci for their valuable contribution 
to the construction of the calorimeter prototype.
We thank Prof. S.C.C. Ting and the L3 collaboration for their cooperation in crystals collection.

This work is partly supported by the project PGR-226 of the Italian Ministry of Foreign Affairs and 
International Cooperation (MAECI), CUP I86D16000060005.

\end{document}